\documentclass{moriond}

\def\Journal#1#2#3#4{{#1} {\bf #2}, #3 (#4)}

\def\CQG{{\em Class. Quantum Grav.}}

\def\PRX{{\em Phys. Rev.} X}

\def\be{\begin{equation}}
\def\ee{\end{equation}}
\def\bea{\begin{eqnarray}}
\def\eea{\end{eqnarray}}

\newcommand{\Photo}{\includegraphics[height=35mm]{LigoVirgo14.jpg}}

\begin{document}
\vspace*{4cm}
\title{LIGO-VIRGO DETECTOR CHARACTERIZATION AND DATA QUALITY (DETCHAR): FROM THE O3 PERFORMANCE TO THE O4 PREPARATION}

\author{ N.~ARNAUD}

\address{Laboratoire Irène Joliot-Curie (IJCLab), Université Paris-Saclay, CNRS/IN2P3, 91405 Orsay, France \\ European Gravitational Observatory (EGO), I-56021 Cascina, Pisa, Italy}

\maketitle\abstracts {
Detector characterization and data quality~-- in short "DetChar"~-- activities are key to optimize the
performance of data-taking periods ("runs") and to turn gravitational-wave (GW) candidates into confirmed
events. The LIGO and Virgo DetChar groups are active from the detector to the final analysis and cover
various latencies: online first, to tag the data that search pipelines can analyze in real-time; then,
the quick vetting (few tens of minutes at most) of the open public alerts targeting the broad astronomer
community for follow-up observations; finally, offline work to define the final datasets and the final
lists of GW events to be published and released publicly. \\
These proceedings summarize the LIGO-Virgo DetChar performance during the O3 run (April 2019 - March 2020) and describe
the main improvements and upgrades that are foreseen for the O4 run that should start during Summer 2022
and include a fourth detector: KAGRA.
}

\section{Introduction}

In the last few years~-- and following decades of design, construction, upgrade, data-taking and data-exploitation phases~--.
50 gravitational-wave (GW) events have been detected\cite{GWTC1,GWTC2} by the global network of terrestrial detectors  made
up of LIGO-Hanford, LIGO-Livingston, Virgo and more recently KAGRA. All these discoveries rely on several pillars that range from
the instruments to the joint analysis of the data. These include detector characterization and data quality
studies~-- globally referred to as "DetChar" in the following.

\begin{figure}
\centerline{\includegraphics[width=0.95\linewidth]{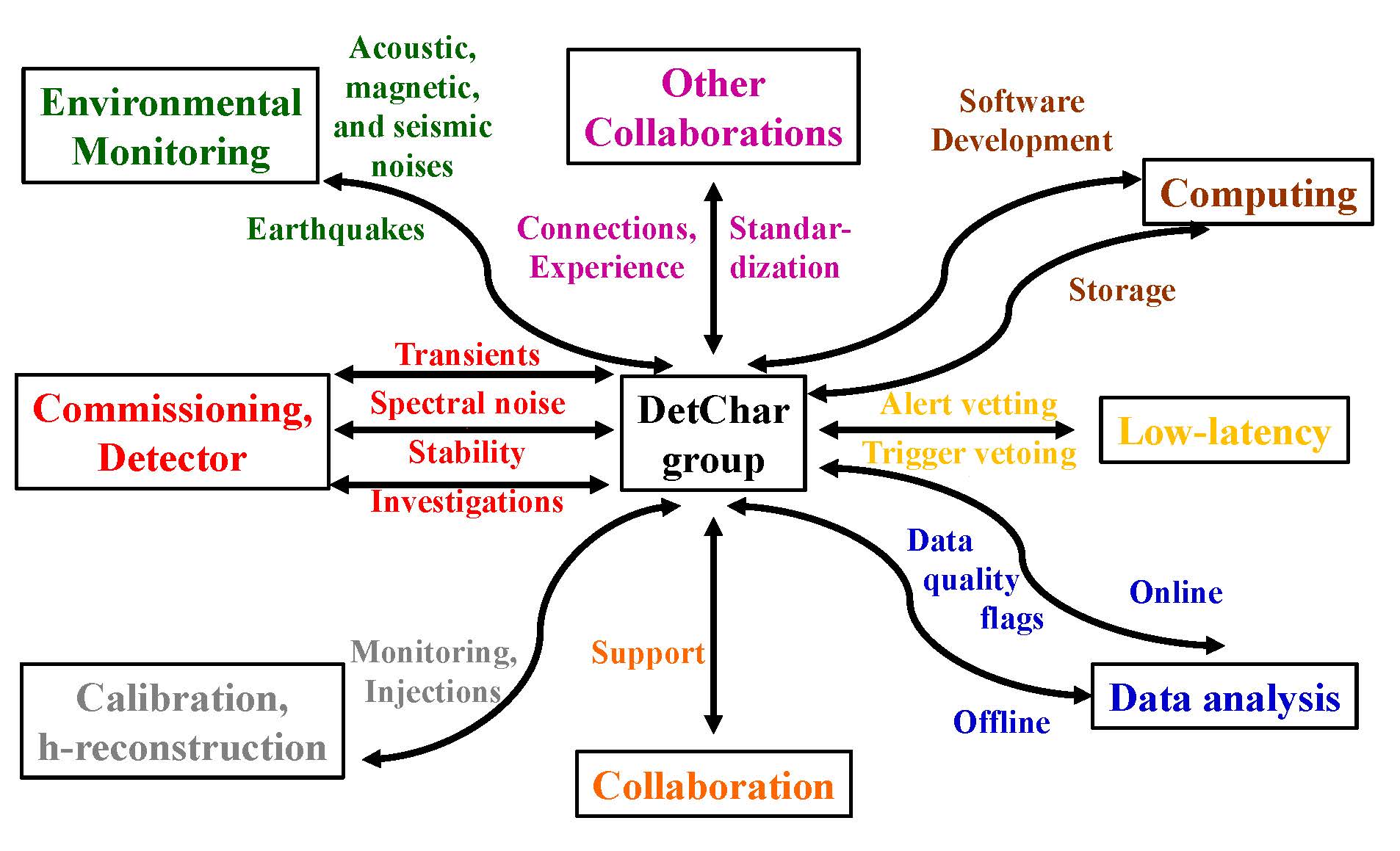}}
\caption[]{Diagram summarizing how DetChar interacts with the main areas of the LIGO-Virgo collaborations. The left side depicts
the instruments while the right one gathers the components that are more connected to data analysis. The top middle box highlights
a key aspect for each individual DetChar group: the cross-collaboration activities that are mandatory to produce joint results and that are very
useful to share experience and develop common frameworks to tackle common problems.}
\label{fig:DetCharFlowchart}
\end{figure}

Fig.~\ref{fig:DetCharFlowchart} shows in broad outline the main DetChar activities and how they are connected to all the other
relevant areas. In addition to supporting detector commissioning phases, DetChar and calibration are strongly coupled to assess
and monitor the quality of the reconstructed GW strain streams. Moreover, an accurate knowledge of the environment around a
given instrument is required to investigate features in its data and to assign them a terrestrial origin. Then, an important part of
the DetChar activities is dedicated to the vetting of GW candidate signals, regardless of whether they have been identified in
low latency by algorithms scanning the live data streams, or found out much later during a re-analyis of the final dataset. The latter
requires beforehand to define such dataset, by identifying and removing bad data quality periods or flagging peculiar data that
require special care to be used. While the validation or the retraction of a public alert demands robust and efficient diagnosis,
able to provide key information with a latency as low as possible. Finally, DetChar also interacts a lot with computing, both to
develop analysis software and to store DetChar products that are made available to the whole collaborations.

With such a rich variety of tasks, it is difficult to review all of them in detail in these proceedings. More information can be found
in~\cite{guide,LIGODetChar,VirgoDetChar} and references therein.

\section{Detector characterization}

GW strain data are usually non-stationary: this is what makes their analysis so complex as one has to continuously estimate the
noise level, so as to be able to detect a real GW signal on top of noise fluctuations. On short timescales, transient bursts of noise, called glitches, are
the main concern. The basic characteristics of a glitch are the following: the frequency range in which it impacts the data, its duration
(or damping time) and its strengh, quantified by its signal-to-noise ratio. During data taking, their rates can vary a lot, depending on
the accuracy with which the detector working point is controlled and on the environment status. For instance, a higher seismic
activity induced by bad weather conditions could stimulate some mechanical resonances that would make a detector component
move more; and if that surface is hit by a parasitic laser beam, that could induce scattered light that would be seen as glitches
in the data.

Therefore, there are many ongoing efforts to classify these glitches into families, using for instance time-frequency representations
called spectrograms. The idea behind that is that similar glitches should have a common origin. Auxiliary data~-- that is all the channels
that are recorded in real time and are expected not to be impacted by the passing of a real GW through the detector~-- are used to
try to find the source of a family of glitches. When a match is found, one can either mitigate the effect of these glitches by knowing when
they occur and rejecting the related GW candidates if there is any, or even make these glitches vanish by fixing their root cause at
the instrument level. Each time a detector gets upgraded, its glitches content changes and analysis must be redone to explore this
new configuration. Strong glitches for which the auxiliary data do not provide information about their origin are hidden by multiplying the
data by a window function (called a "gate") whose weight is one far from the gate, goes smoothly down to zero as the glitch approaches
and is exactly zero when the glitch is active. Such gating preserves the ability to estimate accurately the noise level, at the price of throwing
away some (small) fraction of the data. Although glitches are the main concern for the search for transient GW signals (like the
coalescence of compact binary systems) they also impact the search for continuous signals (from pulsars, stochastic backgrounds, etc.):
those analysis use similar gating procedures to clean the data they analyze.

Glitches can also overlap with real GW signals or be close enough to spoil the estimation of the source parameters. In that case,
sophisticated methods have been developed to excise the glitch without impacting the remainder of the data. This intervention
has been crucial to achieve an accurate sky localization of GW170817 and has become a regular step of the event validation
process during the O3 run. 

Search for continuous GW signal is also impacted by spectral lines, that is frequencies at which the noise is (much) larger than in the
neighboring  frequency range. These lines can be permanent (like the harmonics of the mains) or intermittent and their characteristics
(peak frequency, frequency range, amplitude) can also vary in time. Like for glitch studies, auxiliary data are very useful to
pinpoint the origin of a spectral line. Hundreds of these lines are worth investigating in the spectrum of GW detectors: many of them
are strong enough to be directly visible in the data while others only get discovered during offline analysis looking for continuous GWs.

\section{Data quality}

The third LIGO-Virgo observing run~--"O3"~-- started on April 1st, 2019 and was expected to last 13 months, with a 1-month
commissioning break in the middle. Its first half, "O3a", ended on October 1st, 2019 while the second half, "O3b", started on
November 1st, 2019. Due to the covid-19 pandemic developing worldwide, O3b was ended prematurely on March 27th, 2020,
about a month earlier than expected. O3 is the longest period so far during which three GW detectors have been observing the
cosmos simultaneously.

\begin{figure}
\centerline{\includegraphics[width=0.95\linewidth]{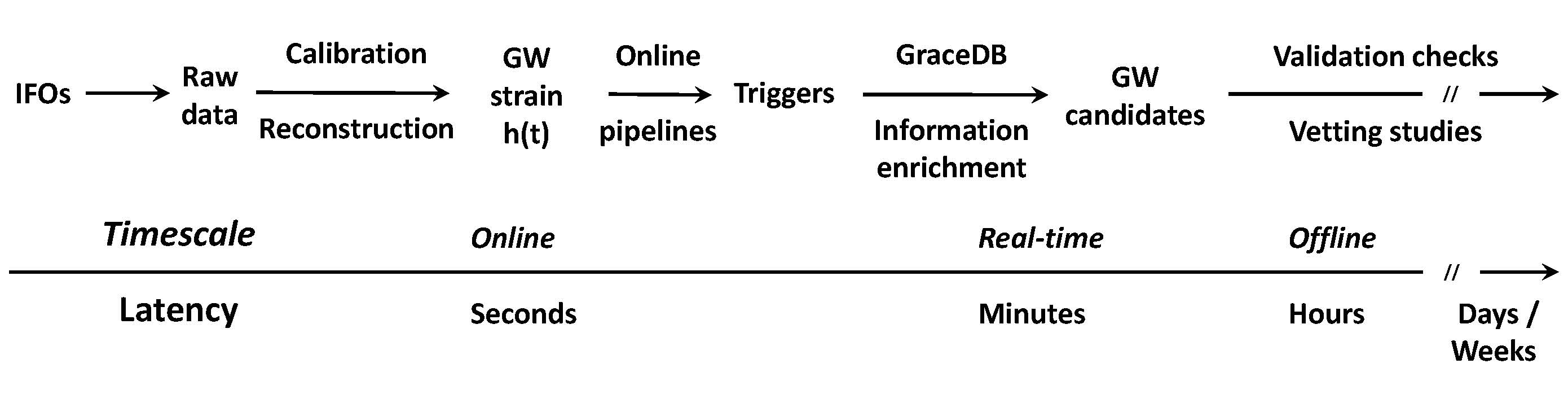}}
\caption[]{Data flow from the detectors~-- "IFOs" (on the left)~-- to the final set of GW detections (on the right).
The bottom part of the diagram shows a
timeline with the corresponding latencies: GW candidate events are generated online with a few seconds latency, vet in real-time
on the timescale of a few minutes and finally confirmed by offline analysis that can take months to be completed and include the
inference of the GW source properties. Each of these steps requires proper inputs from DetChar.}
\label{fig:DataflowDetChar}
\end{figure}

Figure~\ref{fig:DataflowDetChar} shows a simplified timeline of the data flow, from the detectors to the validation of the final
GW events detected. DetChar is present in the three timescales that are highlighted.

\begin{itemize}
\item Instantaneous data quality information is provided to the
algorithms that are scanning the data online, searching for GW candidate signals. This input is twofold: first a veto mode, tagging
the bad data chunks that are not to be analyzed live; and secondly more fine-grained information to help eliminating or
down-ranking triggers that are likely of terrestrial origin according to auxiliary data processed parallel to the GW strain data streams.
\item Then, the public alert candidates identified online need to be vet in real time, that is with a latency of a few tens of minutes at most.
To meet this requirement, DetChar has developed a dedicated framework, called data quality reports~-- in short "DQRs". A DQR is a set
of data quality checks that are triggered automatically when a GW candidate significant-enough is found in the data. These checks
run in parallel and each time one completes, its results are made available immediately on a webpage browsed by the members of
the LIGO-Virgo "rapid response team" who met without warning during O3 to decide whether a public alert is confirmed or needs to be retracted.
All in all, there were 80 public alerts during the O3 run, out of which 24 were retracted (8 during O3a and 16 during O3b).
\item Finally, for offline analysis, Detchar defines the final dataset by extending the data quality analysis that could be run online with low
latency. In addition, it also provides a set of additional recommendations that analysis can choose to use or not, depending on whether
they are sensitive to the particular backgrounds characterized by DetChar.
\end{itemize}

\section{Outlook}

While there are still strong uncertainties linked to the covid-19 pandemic, the fourth LIGO-Virgo-KAGRA observing run~--
"O4"~-- should start not earlier than July 2022 and last at least a year. The O3-O4 shutdown is exploited by all collaborations
to upgrade their instruments (in order to achieve better sensitivities and higher duty cycles) and improve the data analysis
methods. DetChar is involved in both fronts and a key player of the global effort that is ongoing to coordinate as widely
as possible the preparation of the O4 run. This future data-taking period is expected to provide a few times more events
than the O3 run: therefore, in addition to reinforcing the procedures and software tools already in use, DetChar is focusing
on automating its different frameworks and to reduce their latencies wherever possible.

\section*{Acknowledgments}

The author would like to thank his colleagues from the Virgo, LIGO and KAGRA DetChar groups for all their
contributions to detector characterization and data quality. While these activities start at the individual
detector level, the results achieved are truly due to joint cross-collaboration works that these short proceedings
are not doing justice to well-enough.

\end{document}